\documentclass[showpacs,twocolumn]{revtex4}
\usepackage{graphicx}
\usepackage{bm}
\usepackage{graphicx}
\usepackage{amsmath}
\usepackage{mathrsfs}
\usepackage{ulem}
\usepackage{color}

\begin{document}

\title{ Searching for $\Xi_{cc}^+$ in Relativistic Heavy Ion Collisions }
\author{Jiaxin Zhao, Hang He, and Pengfei Zhuang}
\address{Physics Department, Tsinghua University and Collaborative Innovation Center of Quantum Matter, Beijing 100084, China}
\date{\today}

\begin{abstract}
We study the doubly charmed baryon $\Xi_{cc}^+$ in high energy nuclear collisions. We solve the three-body Schr\"odinger equation with relativistic correction and calculate the $\Xi_{cc}^+$ yield and transverse momentum distribution via coalescence mechanism. For $\Xi_{cc}^+$ production in central Pb + Pb collisions at LHC energy, the yield is extremely enhanced, and the production cross section per binary collision is one order of magnitude larger than that in p+p collisions. This indicates that, it is most probable to discover $\Xi_{cc}^+$ in heavy ion collisions and its discovery can be considered as a probe of the quark-gluon plasma formation.
\end{abstract}

\maketitle

%%%%%%%%%%%%%%%%%%%%%%Main body%%%%%%%%%%%%%%%%%%%%%%

The flavor SU(4) quark model predicts 22 charmed baryons~\cite{gellmann}. Searching for them experimentally has been an active direction in the field of high energy physics, but most of them are not yet discovered. The main reason for hardly observing charmed baryons, especially multicharmed baryons, in elementary collisions like p+p and $e^++e^-$ is the rare production of charm quarks in these collisions. For instance, for the production of triply charmed baryon $\Omega_{ccc}$, it requires at least three pairs of charm quarks with small relative momenta in an event. This is very difficult even at LHC energy. In high energy nuclear collisions, however, there are plenty of off-diagonal charm quarks in the created fireball. For instance, the $c\bar c$ pair number can reach 10 in Au + Au collisions at RHIC energy and 100 in Pb+Pb collisions at LHC energy~\cite{gaval}. These uncorrelated charm quarks can be combined to form charmed hadrons via statistics. Obviously, the combination will largely enhance the yield of multicharmed baryons.

The production of charmed hadrons in high energy nuclear collisions is closely related to the Quark-Gluon Plasma (QGP) formation in the early stage of the collisions. Due to the rapid expansion of the colliding system, the temperature of the formed fireball drops down fast, and one cannot directly see the QGP in the final state of the collisions. In 1986, Matsui and Satz pointed out the $J/\psi$ suppression as a signature of the QGP formation~\cite{matsui}. Considering the combination of $c$ and $\bar c$, the $J/\psi$ regeneration~\cite{pbm} in QGP and its competition~\cite{thews,rapp,yan} with the initial production explain well the experimental data of charmonium yield and transverse momentum distributions at RHIC and LHC energies. Extending to the production of multicharmed baryons, the largely enhanced production cross section via the combination of charm quarks in QGP makes them as a unique probe of the new state of matter~\cite{he}.

In this Letter we investigate the production of doubly charmed baryon $\Xi_{cc}^+$ in high energy nuclear collisions. The experimental search for $\Xi_{cc}^+$ lasts for decades. The SELEX collaboration~\cite{selex} claimed the observation of $\Xi_{cc}^+$ in 2003, but the FOCUS~\cite{focus}, Belle~\cite{belle}, BaBar~\cite{babar} and LHCb~\cite{lhcb} collaborations failed to reproduce the results. In comparison with triply charmed $\Omega_{ccc}$~\cite{he}, the production probability of doubly charmed baryons should be much larger, and the decay modes of $\Xi_{cc}^+$ are already widely discussed theoretically and experimentally~\cite{selex,focus,belle,babar,lhcb,onis}. We will first solve the three-body Schr\"odinger equation, including relativistic correction, to obtain the wave function and Wigner function of the ground bound state of three quarks $ccq$, and then calculate the $\Xi_{cc}^+$ yield and transverse momentum distribution in heavy ion collisions through the coalescence mechanism on the hypersurface of deconfinement phase transition determined by hydrodynamics.

We employ the Schr\"odinger equation to describe the bound state of three quarks $ccq$,
\begin{eqnarray}
\label{sch}
&& \hat H \Psi({\bf r}_1,{\bf r}_2,{\bf r}_3) = E_T \Psi({\bf r}_1,{\bf r}_2,{\bf r}_3),\nonumber\\
&& \hat H =\sum_{i=1}^3 {{\bf \hat p}_i^2\over 2m_i}+ V({\bf r}_1,{\bf r}_2,{\bf r}_3)
\end{eqnarray}
with the total energy $E_T$ and wave function $\Psi$. To be specific, we choose the index $i=1,2$ for the two charm quarks and $i=3$ for the light quark. As a usually used approximation~\cite{nielsen}, we neglect the three-body interaction and express the potential as a sum of pair interactions, $V=\sum_{i<j}v({\bf r}_i,{\bf r}_j)$. According to the leading order QCD, the diquark potential is only one half of the quark-antiquark potential. If one assumes such a relation in the case of strong coupling and take the Cornell potential, there is $v({\bf r}_i,{\bf r}_j)=\left(-\alpha/|{\bf r}_i-{\bf r}_j|+\sigma|{\bf r}_i-{\bf r}_j|\right)/2$, where $\alpha=\pi/12$ and $\sigma=0.2$ GeV$^2$ are coupling parameters of the potential which together with the quark masses $m_c=1.25$ GeV and $m_q=0.3$ GeV reproduce well the $D, J/\psi$ and $\psi'$ masses in vacuum. In hot and dense medium, the interaction among quarks should be weakened. From the lattice calculation~\cite{lattice}, however, the $J/\psi$ spectral function is clearly broadened only when the temperature is much higher than the critical temperature $T_c$. Therefore, we can still take the Cornell potential at the coalescence which happens at $T_c$.

\begin{figure}[htb]
{\includegraphics[width=0.3\textwidth]{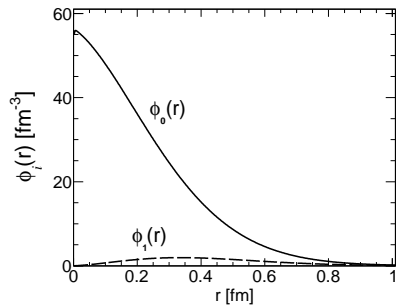}
\caption{The first two hyperradial components $\phi_0(r)$ and $\phi_1(r)$ for the relative wave function of $\Xi_{cc}^+$. }
\label{fig1}}
\end{figure}

We take the hyperspherical method~\cite{nielsen} to solve the three-body Schr\"odinger equation. Introducing the global coordinate ${\bf R}$ and relative coordinates ${\bf r}_x$ and ${\bf r}_y$ through the transformation ${\bf R}=(m_c{\bf r}_1+m_c{\bf r}_2+m_q{\bf r}_3)/M$, ${\bf r}_x=\sqrt{m_c/(2\mu)}({\bf r}_1-{\bf r}_2)$ and ${\bf r}_y=\sqrt{2m_cm_q/(\mu M)}\left({\bf r}_3 -({\bf r}_1+{\bf r}_2)/2\right)$ with the total mass $M=2m_c+m_q$ and an arbitrary parameter $\mu$ with mass dimension which disappears in the end and its value does not affect the result. With the new coordinates, the motion of the three-quark bound state can be factorized into the global motion and relative motion, $\Psi({\bf R},{\bf r}_x,{\bf r}_y)=\Theta({\bf R})\Phi({\bf r}_x,{\bf r}_y)$. By rewriting the amplitudes of ${\bf r}_x$ and ${\bf r}_y$ in terms of the hyperradius $r=\sqrt{r_x^2+r_y^2}$ and hyperpolar angle $\alpha=\arctan (r_x/r_y)$ and constructing the 6-dimension relative motion space $(r,\Omega)=(r, \alpha, \theta_x, \varphi_x, \theta_y, \varphi_y)$ where $\theta_x, \varphi_x$ and $\theta_y, \varphi_y$ are azimuthal angles of ${\bf r}_x$ and ${\bf r}_y$, the kinetic energy in the center of mass frame can be separated into a radial part and an angular part with corresponding eigen states ${\cal R}(r)$ and $Y_\kappa(\Omega)$~\cite{hyper}, where $\kappa$ stands for the 5 quantum numbers $(k,l,m,l_x,l_y)$ in the triplet-singlet representation. However, the potential $V(r,\Omega)$ depends on both the hyperradius $r$ and the 5 angles, the relative motion cannot be factorized into a radial part and an angular part. Considering that the eigen states $Y_\kappa(\Omega)$ constitute a complete set, we express the relative wave function as a linear combination of them,
\begin{equation}
\label{exp}
\Phi(r,\Omega)=\sum_\kappa \phi_\kappa(r) Y_\kappa(\Omega).
\end{equation}
Substituting the expansion into the Schr\"odinger equation and taking the orthonormal relations for $Y_\kappa(\Omega)$~\cite{hyper}, we obtain the coupled ordinary differential equations for the hyperradial wave functions $\phi_\kappa(r)$,
\begin{equation}
\left[\frac{\partial^2 }{\partial r^2}+\frac{5}{r}\frac{\partial }{\partial r}-\frac{k(k+4)}{r^2}+\epsilon\right]\phi_\kappa
=\sum_{\kappa'} V_{\kappa\kappa'}\phi_{\kappa'},
\end{equation}
where $V_{\kappa\kappa'}$ is the potential element
\begin{equation}
V_{\kappa\kappa'}(r)=\int Y_\kappa^*(\Omega) V(r,\Omega)Y_{\kappa'}(\Omega) d\Omega
\end{equation}
with $d\Omega=\sin^2\alpha\cos^2\alpha\sin\theta_x\sin\theta_yd\alpha d\theta_x d\theta_y d\varphi_x d\varphi_y$ and binding energy $\epsilon$. For the ground bound state $\Xi_{cc}^+$, we take only the first two hyperspherical harmonic functions $Y_0=1/\pi^{3/2}$ and $Y_1=2\cos(2\alpha)/\pi^{3/2}$,
\begin{equation}
\label{trun}
\Phi(r,\Omega)=\phi_0(r) Y_0(\Omega)+\phi_1(r) Y_1(\Omega).
\end{equation}
This leads to two coupled equations for the first two hyperradial components $\phi_0$ and $\phi_1$,
\begin{eqnarray}
\label{radial}
&& \left[-\frac{\partial^2 }{\partial r^2}-\frac{5}{r}\frac{\partial }{\partial r}+V_{00}-\epsilon\right]\phi_0+V_{01}\phi_1=0,\nonumber\\
&& \left[-\frac{\partial^2 }{\partial r^2}-\frac{5}{r}\frac{\partial }{\partial r}+\frac{12}{r^2}+V_{11}-\epsilon\right]\phi_1+V_{10}\phi_0=0.
\end{eqnarray}
The two ordinary differential equations can be numerically solved by using the Inverse Power Method~\cite{inverse}, and the radial components $\phi_0(r)$ and $\phi_1(r)$ for the ground state $\Xi_{cc}^+$ are shown in Fig.\ref{fig1}. The second component $\phi_1$ is much smaller than the first component $\phi_0$ at any $r$, indicating that the truncation (\ref{trun}) is good enough for $\Xi_{cc}^+$.
%%%%%%%%%%%%%%%%%%%%%%second picture%%%%%%%%%%%%%%%%%
%%%%%%%%%%%%%%%%%%%%%%%%%%%%%%%%%%%%%%%%%%%%%%%%%%%%%
%%%%%%%%%%%%%%%%%%%%%%%%%%%%%%%%%%%%%%%%%%%%%%%%
\begin{figure}[htb]
{\includegraphics[width=0.3\textwidth]{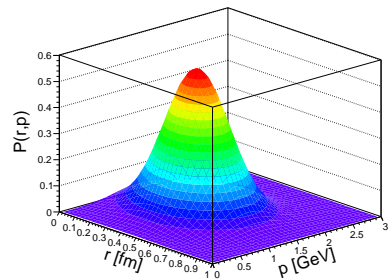}
\caption{The probability $P(r, p)$ to find the three quarks $ccq$ in the ground bound state $\Xi_{cc}^+$ with relative distance $r$ and relative momentum $p$. }
\label{fig2}}
\end{figure}
%%%%%%%%%%%%%%%%%%%%%%%%%%%%%%%%%%%%%%%%%%%%%%%

For $\Xi_{cc}^+$, it is necessary to consider the relativistic correction to the light quark motion. Taking the correction $\Delta\hat H=-\sum_i\hat{\bf p}_i^4/(8m_i^3)$ to the Hamiltonian, the resulted correction to the binding energy and relative wave function of $\Xi_{cc}^+$ follow the perturbation formula,
\begin{eqnarray}
&& \Delta\epsilon_0 = \langle\Phi_0|\Delta\hat H|\Phi_0\rangle + \sum_n {|\langle\Phi_n|\Delta\hat H|\Phi_0\rangle |^2\over \epsilon_0-\epsilon_n}, \nonumber  \\
&& \Delta\Phi_0(r,\Omega) = \sum_n {\langle\Phi_n|\Delta\hat H|\Phi_0\rangle\over \epsilon_0-\epsilon_n}\Phi_n,
\end{eqnarray}
where the sum is over the excited states determined by the radial equations (\ref{radial}). With the correction, we obtain the $\Xi_{cc}^+$ mass and average radius
\begin{eqnarray}
&& m_{\Xi_{cc}^+}=2m_c+m_q+\epsilon_0+\Delta\epsilon_0=3.584\ \text{GeV},\nonumber\\
&& \langle r\rangle_{\Xi_{cc}^+} ={\int drd\Omega r^6\left|\Phi_0+\Delta\Phi_0\right|^2\over \int drd\Omega r^5\left|\Phi_0+\Delta\Phi_0\right|^2}=0.41\ \text{fm}.
\end{eqnarray}
The mass value agrees well with the previous calculations in MIT bag model~\cite{he2}, QCD sum rule~\cite{wang,kiselev}, potential model~\cite{martynenko,tong,lichtenberg} and lattice QCD~\cite{lewis,flynn}, where the $\Xi_{cc}^+$ mass is in between 3.5 GeV and 3.7 GeV.

If $\Xi_{cc}^+$ is considered as the ground bound state of the quark($q$)-diquark($cc$) system, its motion is described in two steps, the diquark motion and the quark-diquark motion. Both are controlled by a two-body Schr\"odinger equation. We recalculated the $\Xi_{cc}^+$ wave function in this case, it is similar to the one obtained with the hyperspherical method, and the $\Xi_{cc}^+$ mass and average radius are respectively $3.62$ GeV and $0.41$ fm. The similarity between the two methods is due to the large mass difference between the light and heavy quarks.

Using the relativistically corrected wave function, we now construct the $\Xi_{cc}^+$ Wigner function in the center of mass frame,
\begin{equation}
W({\bf r},{\bf p})=\int d^6 {\bf y}e^{-i{\bf p}\cdot{\bf y}}\Phi({\bf r}+{{\bf y}\over 2})\Phi^*({\bf r}-{{\bf y}\over 2}),
\end{equation}
where ${\bf p}$ is the 6-dimension momentum corresponding to the relative coordinate ${\bf r}$. Taking the first axis of the integrated vector ${\bf y}$ in the direction of ${\bf p}$ and the second axis on the plane constructed by ${\bf p}$ and ${\bf r}$, the Wigner function is largely simplified and depends only on $r, p, \alpha$ and the angle $\theta$ between ${\bf r}$ and ${\bf p}$. By integrating out the angles, we obtain the probability to find the three quarks in the ground bound state $\Xi_{cc}^+$ with relative distance $r$ and relative momentum $p$,
\begin{equation}
P(r,p)={r^5 p^5\over 24\pi}\int W(r,p,\alpha,\theta)\sin^2\alpha\cos^2\alpha\sin^4\theta d\alpha d\theta
\end{equation}
which is shown in Fig.\ref{fig2}. It is close to a double Gaussian distribution $e^{-(r-\langle r\rangle)^2/\sigma_r^2}e^{-(p-\langle p\rangle)^2/\sigma_p^2}$ with the most probable position at $(\langle r\rangle, \langle p\rangle)=$(0.41 fm, 1.4 GeV) and the standard deviation $(\sigma_r^2, \sigma_p^2)=(\langle (r-\langle r\rangle)^2\rangle, \langle (p-\langle p\rangle)^2\rangle)=( (0.16\ \text{fm})^2, (0.39\ \text{GeV})^2)$.

We now calculate the $\Xi_{cc}^+$ production via coalescence mechanism in high energy nuclear collisions. The coalescence mechanism~\cite{fries} has been successfully used to describe the light hadron production, especially the quark number scaling of the elliptic flow~\cite{molnar} and the enhancement of the baryon to meson ratio~\cite{fries2,greco}. In coalescence models, the change in the constituent distribution before and after the coalescence process is required to be small, namely the number of constituents involved in the coalescence must be small compared with the total particle number of the system. In this sense, the coalescence mechanism is more suitable for the production of rare particles like $\Xi_{cc}^+$. The coalescence probability, namely the Wigner function, is usually parameterized as a Gaussian distribution~\cite{fries,chen} and the width is fixed by fitting the data. For multicharmed baryons, however, there are currently no data, and an adjustable coalescence probability will lose the prediction power of the calculation. For $\Xi_{cc}^+$ we use the above calculated Wigner function $W$ as the coalescence probability.

The QGP created in the early stage of a heavy ion collision is very close to an ideal fluid and its space-time evolution is controlled by hydrodynamic equations $\partial_\mu T^{\mu\nu}=0$ with $T^{\mu\nu}$ being the energy-momentum tensor. The initial condition at time $\tau_0=0.6$ fm/c is determined by the colliding energy and nuclear geometry, which leads to a maximum initial temperature $T_0=484$ MeV in central Pb+Pb collisions at $\sqrt{s_{NN}}=2.76$ TeV~\cite{hirano}. To close the evolution equations, we take the equation of state of the hot medium with a first order phase transition between the ideal QGP and hadron gas at critical temperature $T_c=165$ MeV. By solving the hydrodynamic equations, one obtains the local temperature $T(x)$ and fluid velocity $u_\mu(x)$ which will be used in the coalescence.

The coalescence happens on the hadronization hypersurface $\sigma_\mu(R)$, and the 4D coordinates $R_\mu=(t,{\bf R})$ on the hypersurface is constrained by the critical temperature $T(R_\mu)=T_c$ which leads to the coalescence time $t=t(T_c, {\bf R})$. The observed momentum distribution of $\Xi_{cc}^+$ via coalescence mechanism can be calculated from the Wigner function~\cite{fries},
\begin{eqnarray}
\label{coa}
\frac{dN}{d^2{\bf P}_Td\eta} &=& C\int {\frac{P^\mu d\sigma_\mu(R)}{(2\pi)^3}{\frac{d^4 r_x d^4 r_y d^4 p_x d^4 p_y}{(2\pi)^6}}}  \\
&&\times F(\tilde r_1, \tilde r_2, \tilde r_3, \tilde p_1, \tilde p_2, \tilde p_3)W( r_x, r_y, p_x, p_y),\nonumber
\end{eqnarray}
where $P_\mu =(P_0,{\bf P})$ is the 4D $\Xi_{cc}^+$ momentum with energy $P_0=\sqrt{m_{\Xi_{cc}^+}^2+{\bf P}^2}$ and 3D momentum ${\bf P}=({\bf P}_T,P_z=P_0\sinh\eta)$, corresponding to the coordinate ${\bf R}$, the constant $C=1/18$ comes from the intrinsic symmetry, and $F$ is the distribution function of the three quarks in phase space. Remember that the Wigner function obtained above is derived in the center of mass frame of $\Xi_{cc}^+$ and the $\Xi_{cc}^+$ moves with a 4-velocity $v_\mu=P_\mu/m_{\Xi_{cc}^+}$ in the laboratory frame, the coordinates $\tilde r_i$ and momenta $\tilde p_i$ in the quark distribution function $F$ and $r_i$ and $p_i$ in the Wigner function $W$ are related to each other via a Lorentz transformation~\cite{he}.

The three quark distribution $F$ can be factorized as $F(\tilde r_1, \tilde r_2, \tilde r_3, \tilde p_1, \tilde p_2, \tilde p_3) = S f_c(\tilde r_1, \tilde p_1) f_c(\tilde r_2, \tilde p_2) f_q(\tilde r_3, \tilde p_3)$, where the constant $S=1/2$ counts the symmetry of the two charm quarks. The light quark motion in QGP is controlled by the Fermi distribution $f_q(\tilde r_3, \tilde p_3)=N_qf(\tilde r_3, \tilde p_3)=N_q/(e^{u_\mu \tilde p_3^\mu/T}+1)$ with the degeneracy factor $N_q=6$ and local velocity $u^\mu(\tilde r_3)$ and temperature $T(\tilde r_3)$ of the fluid. The charm quarks are produced through initial hard processes (their regeneration in the QGP at $\sqrt{s_{NN}}=2.76$ TeV is very small and can safely be neglected~\cite{zhou}) and then interact with the hot medium. The single charm quark distribution $f_c$ is in principle between the pQCD distribution with weak interaction and equilibrium distribution with strong interaction. From the experimental data at LHC~\cite{alice}, the observed large quench factor and elliptic flow for charmed mesons indicate that charm quarks are almost thermalized with the medium. Therefore, one can take, as a good approximation, a kinetically equilibrated distribution $f_c(\tilde r_i, \tilde p_i) = \rho_c(\tilde r_i)N_c(\tilde r_i)f(\tilde r_i, \tilde p_i)$, where $N_c(\tilde r_i)$ is the normalization factor of the Fermi distribution. Different from light quarks, charm quarks are not chemically equilibrated in QGP~\cite{kostyuk}, and the space-time evolution of the number density $\rho_c$ is governed by the conservation law during the expansion~\cite{zhou},
\begin{equation}
\partial_\mu \left(\rho_c(\tilde r_i)u^\mu\right)=0
\end{equation}
with the initial number density $\rho_c(\tau_0, \tilde {\bf r}_i)$ determined by the colliding nuclear geometry and the cross section $\sigma_{pp}^{c\bar c}$ of charm quark pair production in p+p collisions.

\begin{figure}[htb]
{\includegraphics[width=0.3\textwidth]{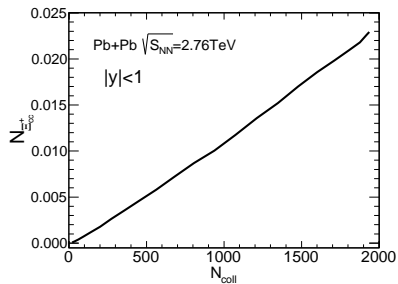}
\caption{The $\Xi_{cc}^+$ yield as a function of the number of binary collisions $N_{coll}$ in Pb+Pb collisions at $|y|<1$ and $\sqrt{s_{NN}}=2.76$ TeV.}
\label{fig3}}
\end{figure}

We can now calculate the $\Xi_{cc}^+$ yield and transverse momentum distribution through the coalescence approach (\ref{coa}) in heavy ion collisions. The yield as a function of the number of binary collisions $N_{coll}$ at middle rapidity $|y| < 1$ in Pb+Pb collisions at $\sqrt{s_{NN}}=2.76$ TeV is shown in Fig.\ref{fig3}. In the calculation we have taken the charm quark cross section $d\sigma_{pp}^{c\bar c}/dy =0.7$ mb~\cite{alice2}. If we consider a homogeneous fireball with volume $V$ and a momentum independent coalescence probability, the $\Xi_{cc}^+$ yield can be estimated as $N_{\Xi_{cc}^+}\sim (N_c/V)^2V=N_c^2/V\sim N_{coll}$, where we have assumed that both the charm quark number $N_c$ and the fireball volume $V$ are proportional to $N_{coll}$. This explains the approximate linear increase in Fig.\ref{fig3}.
%%%%%%%%%%%%%%%%%%%%%%%%yield of \xi%%%%%%%%%%%%%%%%%%%%%%
%%%%%%%%%%%%%%%%%%%%%%%%%%%%%%%%%%%%%%%%%%%%%%%%%%%%%%
%%%%%%%%%%%%%%%%%%%%Pt spectrum%%%%%%%%%%%%%%%%%%%%%%%%%%%%
\begin{figure}[htb]
{\includegraphics[width=0.3\textwidth]{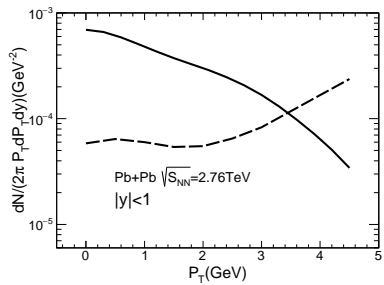}
\caption{The $\Xi_{cc}^+$ transverse momentum distribution (solid line) and the ratio of $\Xi_{cc}^+$ to $J/\psi$ (multiplied by $10^{-2}$, dashed line) in central Pb+Pb collisions at $|y|<1$ and $\sqrt{s_{NN}}=2.76$ TeV.}
\label{fig4}}
\end{figure}
%%%%%%%%%%%%%%%%%%%%%%%%%%%%%%%%%%%%%%%%%%%%%%%%%%%%%%%%

To compare with the $\Xi_{cc}^+$ yield in p+p collisions, we introduce the total production cross section $\tilde\sigma_{AA}^{\Xi_{cc}^+}=N_{\Xi_{cc}^+}/\Delta y\sigma_{pp}$ and the effective cross section per binary collision $\tilde\sigma_{pp}^{\Xi_{cc}^+}=\tilde\sigma_{AA}^{\Xi_{cc}^+}/N_{coll}$ with the inelastic proton cross section $\sigma_{pp}=62$ mb. For most central Pb+Pb collisions with $N_{coll}=2000$, we have $\tilde\sigma_{AA}^{\Xi_{cc}^+}=8.24\cdot 10^5$ nb and $\tilde\sigma_{pp}^{\Xi_{cc}^+}=412$ nb. In comparison with the production cross section $\sigma_{pp}^{\Xi_{cc}^+}=61$ nb in elementary p+p collisions~\cite{berezhony} at $\sqrt{s_{NN}}=14$ TeV and considering the difference in the colliding energy between A+A and p+p collisions, $\tilde\sigma_{AA}^{\Xi_{cc}^+}$ and $\tilde\sigma_{pp}^{\Xi_{cc}^+}$ are, respectively, at least $4$ orders and $1$ order of magnitude larger than $\sigma_{pp}^{\Xi_{cc}^+}$.

In heavy ion collisions, transverse motion is developed during the dynamical evolution of the system and is sensitive to the hot medium properties. In order to understand the $\Xi_{cc}^+$ production mechanism and extract the properties of the medium, we calculated the $\Xi_{cc}^+$ transverse momentum distribution shown in Fig.\ref{fig4}. Due to the statistical law, the $\Xi_{cc}^+$s produced via coalescence mechanism are mainly distributed at low momentum with averaged momentum $\langle P_T\rangle \simeq 2$ GeV which should be much smaller than that through hard processes in p+p collisions. As a consequence of the coalescence mechanism, the feature of the increasing baryon to meson ratio at intermediate and high transverse momentum in heavy ion collisions~\cite{fries2,greco} remains for the $\Xi_{cc}^+$ to $J/\psi$ ratio, see the dashed line in Fig.\ref{fig4}.

We briefly discuss the decay modes of $\Xi_{cc}^+$ which are closely related to its experimental discovery. Since $\Xi_{cc}^+$ is the ground state of the baryons with three quarks $ccq$, its decay is via weak interaction. Possible decay modes include~\cite{selex,focus,belle,babar,lhcb,onis} $\Xi_{cc}^+\to (\Lambda_c^+\to pK^-\pi^+) K^- \pi^+$, $\Xi_{cc}^+\to D^0 p K^- \pi^+$, $\Xi_{cc}^+ \to D^+ p K^-$, $\Xi_{cc}^+ \to \Xi_c^+ \pi^+ \pi^-$ and $\Xi_{cc}^+\to \Xi_c^0 \pi^+$. The lifetime calculated via optical theorem is in between 110 and 250 fs~\cite{kiselev2,guberina}. The short lifetime is probably a challenge to the experimental measurement of $\Xi_{cc}^+$.

In summary, we studied the doubly charmed baryon $\Xi_{cc}^+$ in high energy nuclear collisions. We solved the Schr\"odinger equation for the bound state of three quarks $ccq$ with relativistic correction, and calculated the corresponding Wigner function which is the coalescence probability for the three quarks to combine into a $\Xi_{cc}^+$ in phase space. For Pb+Pb collisions at LHC energy, we computed the $\Xi_{cc}^+$ yield and transverse momentum distribution through coalescence. We found that, the $\Xi_{cc}^+$ production is extremely enhanced, and the effective production cross section per binary collision is already one order of magnitude larger than that in p+p collisions. This indicates that, it is most probable to find $\Xi_{cc}^+$ in heavy ion collisions at LHC energy, and its discovery can be taken as a signal of the QGP formation.

\appendix {\bf Acknowledgement}: The work is supported by the NSFC and MOST grant Nos. 11335005, 11575093, 2013CB922000 and 2014CB845400 and Tsinghua University Initiative Scientific Research Program.

\end{document}